\documentclass[traditabstract,onecolumn]{aa}
\usepackage{amsmath}
\usepackage{amssymb}
\usepackage{txfonts}
\usepackage{graphicx}
\usepackage[colorlinks=true, linkcolor=blue, citecolor=blue, urlcolor=blue]{hyperref}
\usepackage{natbib}

\DeclareMathOperator*{\Res}{Res}
\DeclareMathOperator*{\erf}{erf}
\newcommand{\reff}{R_{\text{e}}}

\bibpunct{(}{)}{;}{a}{}{,} 
\pdfoutput=1
\begin{document}

\title{Analytical expressions for the deprojected S\'ersic model\\
  II. General expressions in terms of the Fox H function}
\titlerunning{Analytical expressions for the deprojected S\'ersic model II}

\author{M. Baes and E. Van Hese}
\authorrunning{M. Baes \& E. Van Hese}

\institute{Sterrenkundig Observatorium, Universiteit Gent,
Krijgslaan 281 S9, B-9000 Gent, Belgium\\ \email{maarten.baes@ugent.be}}

\date{\today}

\abstract{The S\'ersic model is the de facto standard to describe the
  surface brightness distribution of hot stellar systems. An important
  inconvenience of this analytical model is that the corresponding
  luminosity density and associated properties cannot be expressed
  using elementary functions or even standard special functions. We
  present a set of compact and elegant analytical expressions for the
  luminosity density, cumulative luminosity and potential for the
  S\'ersic model in terms of the Fox $H$ function for general values
  of the S\'ersic index. Furthermore, we present explicit series
  expansions of these quantities and discuss the asymptotic
  behaviour. Our analysis completes the work of Mazure \& Capelato
  (2002) and Baes \& Gentile (2011) and demonstrates the power of the
  underestimated Fox $H$ function as a tool for analytical work.  }

\keywords{methods: analytical -- galaxies: photometry}

\maketitle

\section{Introduction}
\label{sec:intro}

\defcitealias{2011A&A...525A.136B}{BG11}

The \citet{1968adga.book.....S} surface brightness profile, has become
the preferred model to describe the surface brightness profile of
early-type galaxies and the bulges of spiral galaxies
\citep[e.g.][]{1988MNRAS.232..239D, 1993MNRAS.265.1013C,
  1994MNRAS.271..523D, 1994ApJS...93..397C, 1995MNRAS.275..874A,
  1997A&A...321..111P, 2001A&A...368...16M, 2003AJ....125.2936G,
  2006MNRAS.371....2A, 2009MNRAS.393.1531G}. Many analytical
properties of this model have been discussed in the literature
\citep[e.g.][]{1991A&A...249...99C, 1997A&A...321..724C,
  1999A&A...352..447C, 2001MNRAS.326..869T, 2002A&A...383..384M,
  2004A&A...415..839C, 2005PASA...22..118G, 2007JCAP...07..006E,
  2011A&A...525A.136B}. 

An important inconvenience of the S\'ersic model is that its
deprojected luminosity density, i.e.\ the spatial 3D luminosity
density $\nu(r)$ that projects on the plane of the sky to the S\'ersic
surface brightness profile, cannot be expressed using elementary
functions or even in terms of standard special functions. It was long
thought that no analytical expression could be obtained. Quite
unexpectedly, \citet{2002A&A...383..384M} came up with an analytical
expression for $\nu(r)$ in terms of the Meijer $G$ function for all
integer S\'ersic indices $m$. \citet[][hereafter
\citetalias{2011A&A...525A.136B}]{2011A&A...525A.136B} took this
analysis one step further and showed that the deprojection of the
S\'ersic surface brightness profile for general values of $m$ can be
solved elegantly using Mellin integral transforms and gives rise to a
Mellin-Barnes integral. The result is that the S\'ersic luminosity
density can be written compactly in terms of a Fox $H$ function, which
reduces to a Meijer $G$ function for all rational values of
$m$. \citetalias{2011A&A...525A.136B} also calculated analytically a
number of additional properties of the S\'ersic model for rational
$m$, including the asymptotic expansion of the luminosity density at
small and large radii, the cumulative light profile and the
gravitational potential. This was possible thanks to the many
analytical properties of the Meijer $G$ function.

The goal of this paper is two-fold. Foremost, we extend and complete
the analysis presented in \citet{2002A&A...383..384M} and
\citetalias{2011A&A...525A.136B}: we will provide compact and elegant
expressions for the density, potential and luminosity profiles in
terms of the general Fox $H$ function, which are valid for {\em{all}}
values of the S\'ersic index $m$ rather than just for integer or
rational $m$. We also present a completely general series expansion of
these functions that enables both a numerical evaluation and a
straightforward analytical study of the asymptotic behaviour. Besides
providing these useful characteristics on one of the most used models
in extragalactic astronomy, our work also has a secondary objective,
namely to demonstrate the power of the Fox $H$ function for analytical
work. The Fox $H$ function is, to use an understatement, not the most
mainstream special function: it is even not contained in the standard
works on special functions such as \citet{1965tisp.book.....G} or the
Wolfram Functions Site. We feel that this is not justified: it is in
fact a very elegant and powerful tool for analytical work and it is
becoming more and more used in mathematics and applied sciences,
including physics, biology, engineering and earth sciences. It is also
gradually being used in astrophysics, e.g.\ for solar and stellar
structure models, fractional reaction-diffusion equations and stellar
dynamics \citep{2007BASI...35..681H, 2011arXiv1102.5498H,
  2007A&A...471..419B, 2009ApJ...690.1280V, 2010ApJ...724L.171D}. With
our analysis, we wish to illustrate its useful properties and advocate
its use in theoretical astrophysical research.

In Section~\ref{sec:analytical} we derive compact expressions for the
luminosity density, cumulative luminosity and potential for the
S\'ersic model, and in Section~\ref{sec:series} we deduce detailed
power and logarithmic-power series expressions for these important
quantities. We discuss the asymptotic behaviour of our functions in
Section~\ref{sec:asymptot}, and we sum up our results in
Section~\ref{sec:conclusions}. In Appendix~\ref{appendix} we present
the Fox $H$ function and we discuss some of its properties that were
used for our analysis.

\section{Analytical properties of the S\'ersic model}
\label{sec:analytical}

The S\'ersic model is defined by the intensity profile projected on
the plane of the sky,
\begin{equation}
  I(R)
  =
  I_0\exp\left[-b\left(\frac{R}{\reff}\right)^{1/m}\right].
\label{Sersic}
\end{equation}
The 3D, deprojected luminosity density $\nu(r)$ of a spherically
symmetric system can be recovered from the surface brightness profile
$I(R)$ using the standard deprojection formula
\begin{equation}
  \nu(r)
  =
  -\frac{1}{\pi}
  \int_r^\infty \frac{{\text{d}}I}{{\text{d}}R}(R)\,
  \frac{{\text{d}}R}{\sqrt{R^2-r^2}}.
  \label{deprojection}
\end{equation}
Substituting the S\'ersic profile (\ref{Sersic}) into
(\ref{deprojection}) we obtain an integral that cannot readily be
evaluated using the standard ways or look-up
tables. \citetalias{2011A&A...525A.136B} applied a Mellin integral
transform technique to convert this integral to a Mellin-Barnes
contour integral,
\begin{equation}
 \nu(r)
  =
 \frac{2m\,I_0}{\sqrt{\pi}}\,r^{-1}\,
 \frac{1}{2\pi i}
  \int_{\mathcal{L}} 
  \frac{\Gamma(2m\,x)\,\Gamma\left(\frac12+x\right)}
  {\Gamma(x)}\,
  \left(\frac{b^m r}{\reff}\right)^{-2x}\,
 {\text{d}}x,
\label{nuHint}
\end{equation}
or, given the definition (\ref{defH}) of the Fox $H$ function, to the
compact expression
\begin{equation}
 \nu(r)
  =
 \frac{2m\,I_0\,b^m}{\sqrt{\pi}\,\reff}\,
 u^{-1}\,
 H^{2,0}_{1,2}\!\left[\left.
    \begin{matrix} 
      (0,1)\\
      (0,2m), (\tfrac12,1)
    \end{matrix}
    \,\right|\,
  u^2
 \right],
\label{nuH}
\end{equation}
where we have used the reduced coordinate
\begin{equation}
  u 
  =
  \frac{b^mr}{\reff}.
\end{equation}
\citetalias{2011A&A...525A.136B} converted this general expression (or
actually, a slightly different but equivalent expression) to
expressions in terms of the Meijer $G$ function for integer and
rational values of the S\'ersic index $m$. As a check on these
formulae, they calculated the total luminosity of the S\'ersic model
for rational $m$ by integrating the luminosity density over the entire
space. To obtain the results, they used the integration properties of
the Meijer $G$ function, combined with several applications of Gauss'
multiplication theorem. As a generalization of this result, and as a
nice example of the power of the Fox $H$ function, we calculate the
total luminosity from the general formula (\ref{nuH}), i.e.
\begin{equation}
  L
  =
  4\pi\int_0^\infty \nu(r)\,r^2\,{\text{d}}r
  =
  \frac{4m\sqrt{\pi}\,I_0\,\reff^2}{b^{2m}}
  \int_0^\infty
  H^{2,0}_{1,2}\!\left[\left.
      \begin{matrix} 
        (0,1)\\
        (0,2m), (\tfrac12,1)
      \end{matrix}
      \,\right|\,
    t
  \right]
  {\text{d}}t.
\label{oja}
\end{equation}
To evaluate this integral, one should remind that
equation~(\ref{defH}) defined the Fox $H$ function as an inverse
Mellin transform of a combination of gamma functions. As a result, the
Mellin transform of a Fox $H$ function reads
\begin{equation}
  \int_0^\infty
  H^{m,n}_{p,q}\!\left[\left.
      \begin{matrix} 
        ({\boldsymbol{a}},{\boldsymbol{A}})
        \\
        ({\boldsymbol{b}},{\boldsymbol{B}})
      \end{matrix}
      \,\right|\, z \right] 
  z^{s-1}\,{\text{d}}z
  =
  \frac{\prod_{j=1}^m \Gamma(b_j+B_js) \prod_{j=1}^n \Gamma(1-a_j-A_js)}
  {\prod_{j=m+1}^q \Gamma(1-b_j-B_js) \prod_{j=n+1}^p\Gamma(a_j+A_js)}.  
\end{equation}
Applying this to (\ref{oja}) with $s=1$, we obtain
\begin{equation}
  L
  =
  \frac{4m\sqrt{\pi}\,I_0\,\reff^2}{b^{2m}}\,
  \frac{\Gamma(2m)\,\Gamma(\tfrac32)}{\Gamma(1)}
  =
  \frac{2\pi\,m\,I_0\,\reff^2\,\Gamma(m)}{b^{2m}},
\end{equation}
in agreement with the value obtained by integrating the surface
brightness profile~(\ref{Sersic}) over the plane of the sky
\citep{1991A&A...249...99C}.

From the luminosity density, a number of other important quantities
can be derived, most importantly the cumulative luminosity profile
$L(r)$ and the gravitational potential $\Psi(r)$, 
\begin{gather}
  L(r) 
  =
  4\pi \int_0^r \rho(r')\,r'^2\,{\text{d}}r'.
  \label{defL}
  \\
  \Psi(r)
  =
  G\,\Upsilon
  \int_r^\infty \frac{L(r')\,{\text{d}}r'}{r'^2},
  \label{defPsi}
\end{gather}
where the $\Upsilon$ is the mass-to-light
ratio. \citet{2002A&A...383..384M} and
\citetalias{2011A&A...525A.136B} calculated these quantities for the
S\'ersic model for integer and rational values of the S\'ersic
parameter, respectively, using the integration properties of the
Meijer $G$ function. It is, however, possible to calculate these
properties for general $m$ in an elegant way by directly applying the
integrations on the Mellin-Barnes integral form of the luminosity
density. For the cumulative luminosity profile we find
\begin{align}
  L(r)
  &=
  8m\sqrt{\pi}\,I_0
  \int_0^r
  \left[
    \frac{1}{2\pi i}
    \int_{\mathcal{L}} 
    \frac{\Gamma(2m\,x)\,\Gamma\left(\frac12+x\right)}{\Gamma(x)}\,
    \left(\frac{b^mr'}{\reff}\right)^{-2x}
    {\text{d}}x
  \right]
  r'\,{\text{d}}r'
  \nonumber \\
  &=
  8m\sqrt{\pi}\,I_0\,
  \frac{1}{2\pi i}
  \int_{\mathcal{L}} 
  \frac{\Gamma(2m\,x)\,\Gamma\left(\frac12+x\right)}{\Gamma(x)}\,
  \left(\frac{b^m}{\reff}\right)^{-2x}
  \left[
    \int_0^r r'^{1-2x}\,{\text{d}}r'
  \right]
  {\text{d}}x
  \nonumber \\
  &=
  4m\sqrt{\pi}\,I_0\,r^2
  \frac{1}{2\pi i}
  \int_{\mathcal{L}} 
  \frac{\Gamma(2m\,x)\,\Gamma\left(\frac12+x\right)\,\Gamma(1-x)}
  {\Gamma(x)\,\Gamma(2-x)}\,
  u^{-2x}\,
  {\text{d}}x
  \nonumber \\
  &=
  \frac{4m\sqrt{\pi}\,I_0\,\reff^2}{b^{2m}}\,
  u^2\,
  H^{2,1}_{2,3}\!\left[\left.
    \begin{matrix} 
      (0,1), (0,1)\\
      (0,2m), (\tfrac12,1),(-1,1)
    \end{matrix}
    \,\right|\,
  u^2
 \right].
\label{LH}
\end{align}  
For the gravitational potential we find after a similar calculation
\begin{equation}
  \Psi(r)
  =
  \frac{2m\sqrt{\pi}\,G\,\Upsilon\,I_0\,\reff}{b^m}\,
  u\,
  H^{2,1}_{2,3}\!\left[\left.
    \begin{matrix} 
      (0,1), (0,1)\\
      (0,2m), (-\tfrac12,1),(-1,1)
    \end{matrix}
    \,\right|\,
  u^2
  \right].
\label{PsiH}
\end{equation}  
The formulae~(\ref{nuH}), (\ref{LH}) and (\ref{PsiH}) form a triplet
of formulae that describe three important spatial properties of the
S\'ersic model in a compact way. 

A straightforward way of checking these formulae is to look at the
model that corresponds to $m=\tfrac12$. In this case, all components
of the vectors $\boldsymbol{A}$ and $\boldsymbol{B}$ are equal to one,
such that the Fox $H$ function reduces to a Meijer $G$
function. We find
\begin{gather}
  \nu(r)
  =
  \frac{I_0\sqrt{b}}{\sqrt{\pi}\,\reff}\,
  u^{-1}\,
  G^{2,0}_{1,2}\!\left(\left.
      \begin{matrix} 
        0 \\
        0, \tfrac12
      \end{matrix}
      \,\right|\,
    u^2
  \right)
  = 
  \frac{I_0\sqrt{b}}{\sqrt{\pi}\,\reff}\,
  {\text{e}}^{-u^2},
 \label{nu12}
  \\
  L(r)
  =
  \frac{2\!\sqrt{\pi}\,I_0\,\reff^2}{b}\,
  u^2\,
  G^{2,1}_{2,3}\!\left(\left.
      \begin{matrix} 
        0, 0 \\
        0, \tfrac12, -1
      \end{matrix}
      \,\right|\,
    u^2
  \right)
  =
  \frac{\pi\,I_0\,\reff^2}{b}\, 
  \left[
    \erf u
    -
    \frac{2}{\sqrt{\pi}}\,u\,{\text{e}}^{-u^2}
  \right],
 \label{L12}
  \\
  \Psi(r)
  =
  \frac{\sqrt{\pi}\,G\,\Upsilon\,I_0\,\reff}{\sqrt{b}}\,
  u\,
  G^{2,1}_{2,3}\!\left(\left.
      \begin{matrix} 
        0, 0 \\
        0, -\tfrac12,-1
      \end{matrix}
      \,\right|\,
    u^2
  \right)
  =
\frac{\pi\,G\,\Upsilon\,I_0\,\reff}{\sqrt{b}}\,
 \frac{\erf u}{u}.
 \label{Psi12}
\end{gather}
These expressions can also be derived by substituting the intensity
profile $I(R)=I_0\,{\text{e}}^{-bR^2/\reff^2}$ in the
expressions (\ref{deprojection}), (\ref{defL}) and (\ref{defPsi}) and
directly evaluating the resulting integrals. More generally, one can
check that the formulae~(\ref{nuH}), (\ref{LH}) and (\ref{PsiH})
reduce to the equations (22), (40) and (44) of
\citetalias{2011A&A...525A.136B} for rational values of
$m$.

\section{Explicit series expansions}
\label{sec:series}

While the expressions~(\ref{nuH}), (\ref{LH}) and (\ref{PsiH}) form an
triplet of compact formulae that are useful for analytical work, they
are not readily useful to numerically evaluate the spatial properties
of the S\'ersic model. For rational values of $m$, the Fox $H$
functions reduce to Meijer $G$ functions, and some numerical software
packages have this function now implemented. However, the numerical
evaluation of Meijer $G$ functions with large parameter vectors (which
easily occurs in our case for rational values of $m$, as can be seen
in \citetalias{2011A&A...525A.136B}), proves to be difficult, in
particular in cases where second-order poles are present in the
integrand of the inverse Mellin transform. Moreover, for general
values of $m$, the expressions~(\ref{nuH}), (\ref{LH}) and
(\ref{PsiH}) cannot be written in terms of the Meijer $G$ function or
any other special function, and we are not aware of any
implementations in numerical software that can evaluate general Fox
$H$ functions.

In this section, we derive explicit series expansions for $\nu(r)$,
$L(r)$ and $\Psi(r)$, which both enables the numerical calculation and
again highlights the power of Fox $H$ function as a useful
mathematical tool. The expansions build on the general series
expansion of the Fox $H$ function as power or power-logarithmic series
(details can be found in Appendix~\ref{appendix}). The form of the
series expansion depends on the multiplicity of the poles of the gamma
functions $\Gamma(b_j+B_js)$. For $\nu(r)$ and $L(r)$, the poles of
these gamma functions are found at $-k_1/2m$ and $-1/2-k_2$ with $k_1$
and $k_2$ any natural number. The gamma functions corresponding to the
expression of the potential $\Psi(r)$ contain the same poles with an
additional pole at $1/2$. The good news is that each pole can at most
occur twice, the bad news is that this happens quite often: for all
integer $m$ and rational $m=p/q$ where the denominator $q$ of the
fraction is odd, double poles do occur.


In the case of simple poles of the gamma functions $\Gamma(b_j+B_js)$,
the expansion of the Fox $H$ function is a power series, given by
equation~(\ref{powerseries}). Applied to our case, we find in case $m$
is non-rational or rational with an even denominator, the expansions
\begin{gather}
  \nu(r) 
  = 
  \frac{2m\,I_0\,b^m}{\sqrt{\pi}\,\reff}\,
  \left[
    \sum_{k=1}^{\infty} \frac{\Gamma\left(\tfrac12 - \tfrac{k}{2m}\right)}{\Gamma\left(-\tfrac{k}{2m}\right)}\,
    \frac{(-1)^k}{k!}\frac{u^{k/m-1}}{2m}
    + 
    \sum_{k=0}^{\infty} \frac{\Gamma(-m - 2mk)}{\Gamma\left(-\tfrac12-k\right)}\,
    \frac{(-1)^k}{k!}u^{2k}
  \right],  
  \label{nugenseries}
  \\
  L(r) 
  = 
  \frac{4m\sqrt{\pi}\,I_0\,\reff^2}{b^{2m}}\,
  \left[
    \sum_{k=1}^{\infty} \frac{\Gamma\left(\tfrac12 - \tfrac{k}{2m}\right)}{\Gamma\left(-\tfrac{k}{2m}\right)}\,
    \frac{(-1)^k}{k!}\frac{u^{k/m+2}}{k+2m}
    + 
    \sum_{k=0}^{\infty} \frac{2\,\Gamma(-m - 2mk)}{\Gamma\left(-\tfrac12-k\right)}\,
    \frac{(-1)^k}{k!}\frac{u^{2k+3}}{2k+3}
  \right],  
  \label{massgenseries}
  \\
  \Psi(r) 
  = 
  \frac{2m\sqrt{\pi}\,G\,\Upsilon\,I_0\,\reff}{b^m}\,
  \left[
    \sum_{k=1}^{\infty} \frac{\Gamma\left(-\tfrac12 - \tfrac{k}{2m}\right)}{\Gamma\left(-\tfrac{k}{2m}\right)}\,
    \frac{(-1)^k}{k!}\frac{u^{k/m+1}}{k+2m}
    + 
    \sum_{k=0}^{\infty} \frac{2\,\Gamma(m - 2mk)}{\Gamma\left(\tfrac12-k\right)}\,
    \frac{(-1)^k}{k!}\frac{u^{2k}}{2k+1}
  \right].
  \label{psigenseries}
\end{gather}
Notice that the term $k=0$ in the first sums is omitted, since the
factors $\Gamma(0)$ in the denominator make those terms vanish. In
fact, if $m$ is a rational number $p/q$ with $q$ even, then the terms
in the first sums for which $k=0,p,2p,\ldots$ vanish; if $p=1$, these
first sums vanish completely. A particularly interesting case is
(again) $m=\tfrac12$, where we find
\begin{gather}
  \nu(r)
  = 
  \frac{I_0\sqrt{b}}{\sqrt{\pi}\,\reff}\,
  \sum_{k=0}^{\infty} \frac{(-1)^k}{k!}u^{2k}
  =
  \frac{I_0\sqrt{b}}{\sqrt{\pi}\,\reff}\,
  {\text{e}}^{-u^2},
 \\
  L(r)
  =
  \frac{4\!\sqrt{\pi}\,I_0\,\reff^2}{b}\, 
  \sum_{k=0}^{\infty} 
  \frac{(-1)^k}{k!}\frac{u^{3+2k}}{2k+3}
  =
 \frac{\pi\,I_0\,\reff^2}{b}\, 
  \left[
    \erf u
    -
    \frac{2}{\sqrt{\pi}}\,u\,{\text{e}}^{-u^2}
  \right],
 \\
  \Psi(r)
  =
  \frac{2\!\sqrt{\pi}\,G\,\Upsilon\,I_0\,\reff}{\sqrt{b}}\,
  \sum_{k=0}^{\infty} \frac{(-1)^k}{k!}\frac{u^{2k}}{2k+1}
  =
  \frac{\pi\,G\,\Upsilon\,I_0\,\reff}{\sqrt{b}}\,
  \frac{\erf u}{u},
\end{gather}
in agreement with formulae~(\ref{nu12}), (\ref{L12}) and (\ref{Psi12}).


When the gamma functions $\Gamma(b_j+B_js)$ have multiple poles, the
expansion of the Fox $H$ function is a logarithmic-power series, the
complexity of which increases with increasing multiplicity of the
poles. The full expression for the case where two gamma functions
share poles is given in equation~(\ref{genseries}). For our present
case, this means that we obtain logarithmic-power series expansions
for $\nu(r)$, $L(r)$ and $\Psi(r)$ if the S\'ersic index $m$ is
integer or rational with an odd denominator. 
If we define $k_0=(q+1)/2$, one obtains after quite some algebra
\begin{multline}
  \quad
  \nu(r)
  = 
  \frac{2m\,I_0\,b^m}{\sqrt{\pi}\,\reff}\,
  \left\{\;
    \sum_{\substack{k=1 \\ k\,{\text{mod}}\,p\ne0}}^\infty
    \hspace{-1ex}
    \frac{\Gamma\left(\tfrac12 - \tfrac{k}{2m}\right)}
    {\Gamma\left(-\tfrac{k}{2m}\right)}\,
    \frac{(-1)^k}{k!}\,\frac{u^{k/m-1}}{2m}
    \ + 
    \hspace{-2.5ex}
    \sum_{\substack{k=0 \\ (k+k_0)\,{\text{mod}}\,q\ne0}}^\infty
    \hspace{-2.5ex}
    \frac{\Gamma(-m - 2mk)}{\Gamma\left(-\tfrac12-k\right)}\,
    \frac{(-1)^k}{k!}u^{2k}\right.
  \\[-1ex]
  \left.
    - \  \frac{1}{\sqrt{\pi}}\hspace{-2.5ex}
    \sum_{\substack{k=0 \\ (k+k_0)\,{\text{mod}}\,q=0}}^\infty
    \hspace{-2.5ex}
    \frac{(-1)^{p}}{2m}\frac{(2k+1)!}{(2km + m)!\,k!\,k!}
    \left(\frac{u}{2}\right)^{2k}
    \ 
    \left[ - \ln\left(\frac{u}{2}\right) + \psi(k+1) + m\,\psi(2km + m)
      - \psi(2k+1)\right]
    \vphantom{    
      \sum_{\substack{k=0 \\ k\,{\text{mod}}\,m\ne0}}^\infty
      \frac{\Gamma\left(\tfrac12 - \tfrac{k}{2m}\right)}
      {\Gamma\left(-\tfrac{k}{2m}\right)}
    }
  \right\},
  \quad
  \label{nuratseries}
\end{multline}
\begin{multline}
  \quad
  L(r)
  = 
  \frac{4m\sqrt{\pi}\,I_0\,\reff^2}{b^{2m}}\,
  \left\{\;
    \sum_{\substack{k=1 \\ k\,{\text{mod}}\,p\ne0}}^\infty
    \hspace{-1ex}
    \frac{\Gamma\left(\tfrac12 - \tfrac{k}{2m}\right)}
    {\Gamma\left(-\tfrac{k}{2m}\right)}\,
    \frac{(-1)^k}{k!}\frac{u^{k/m+2}}{k+2m}
    \ + \ 
    \hspace{-2.5ex}
    \sum_{\substack{k=0 \\ (k+k_0)\,{\text{mod}}\,q\ne0}}^\infty
    \hspace{-2.5ex}
    \frac{2\,\Gamma(-m - 2mk)}{\Gamma\left(-\tfrac12-k\right)}\,
    \frac{(-1)^k}{k!}\frac{u^{2k+3}}{2k+3}\right.
  \\[-1ex]
  \left.
    - \  \frac{u^3}{\sqrt{\pi}}
    \hspace{-2.5ex}
    \sum_{\substack{k=0 \\ (k+k_0)\,{\text{mod}}\,q=0}}^\infty
    \hspace{-2.5ex}
    \frac{(-1)^{p}}{2km+3m}\frac{(2k+1)!}{(2km + m)!\,k!\,k!}
    \left(\frac{u}{2}\right)^{2k}
    \
    \left[ - \ln\left(\frac{u}{2}\right) + \frac{1}{2k+3} + \psi(k+1) + m\,\psi(2km + m)
      - \psi(2k+1)\right]
    \vphantom{
      \sum_{\substack{k=0 \\ k\,{\text{mod}}\,m\ne0}}^\infty
      \frac{\Gamma\left(\tfrac12 -
          \tfrac{k}{2m}\right)}{\Gamma\left(-\tfrac{k}{2m}\right)}
    }
  \right\},
  \quad
  \label{massratseries}
\end{multline}
\begin{multline}
  \quad
  \Psi(r) 
  = 
  \frac{2m\sqrt{\pi}\,G\,\Upsilon\,I_0\,\reff}{b^m}
  \left\{\;
    \sum_{\substack{k=1 \\ k\,{\text{mod}}\,p\ne0}}^\infty
    \hspace{-1ex}
    \frac{\Gamma\left(-\tfrac12 - \tfrac{k}{2m}\right)}
    {\Gamma\left(-\tfrac{k}{2m}\right)}\,
    \frac{(-1)^k}{k!}\frac{u^{k/m+1}}{k+2m}
    \ - \ 
    \hspace{-2.5ex}
    \sum_{\substack{k=0 \\ (k+k_0)\,{\text{mod}}\,q\ne0}}^\infty
    \hspace{-2.5ex}
    \frac{2\,\Gamma(-m - 2mk)}{\Gamma\left(-\tfrac12-k\right)}\,
    \frac{(-1)^k}{(k+1)!}\frac{u^{2k+2}}{2k+3}
    \ + \
    \frac{2\Gamma(m)}{\sqrt{\pi}}
  \right.
  \\[-1ex]
  \left.
    + \  \frac{u^2}{\sqrt{\pi}}\hspace{-2.5ex}
    \sum_{\substack{k=0 \\ (k+k_0)\,{\text{mod}}\,q=0}}^\infty
    \hspace{-2.5ex}
    \frac{(-1)^{p}}{2km+3m}\frac{(2k+1)!}{(2km + m)!\,k!\,(k+1)!}
    \left(\frac{u}{2}\right)^{2k}
    \
    \left[ 
      - \ln\left(\frac{u}{2}\right) + \frac{1}{2k+2} + \frac{1}{2k+3} 
      + \psi(k+1) + m\,\psi(2km + m) - \psi(2k+1)
    \right]
    \vphantom{
      \sum_{\substack{k=0 \\ k\,{\text{mod}}\,m\ne0}}^\infty
      \frac{\Gamma\left(-\tfrac12 -
          \tfrac{k}{2m}\right)}{\Gamma\left(-\tfrac{k}{2m}\right)}
    }
  \right\}.
  \label{psiratseries}
\end{multline}
In these expressions, $\psi(s)=\Gamma'(s)/\Gamma(s)$ is the digamma
function.  In $\Psi(r)$, the third term $2\Gamma(m)/\!\sqrt{\pi}$
corresponds with the residue of the pole $\tfrac12$.  Again, the terms
in the first sums for which $k=0,p,2p,\ldots$ vanish, and these first
sums vanish completely if $p=1$. On the other hand, if $q=1$, i.e.\ if
$m$ is an integer value, the second sums vanish, since then the
integrands for $\nu(r)$ and $L(r)$ have no simple poles
$\beta_{2,k_2}$, while for $\Psi(r)$ only the pole $\tfrac12$ remains
as a single pole.
On the crossroad of these two cases we have $m=1$: for this model,
both the first and second sums in the
expansions~(\ref{nuratseries})-(\ref{psiratseries}) vanish completely,
apart from a single term for the potential. Astrophysically, the
S\'ersic model with $m=1$ corresponds to a model with an exponential
surface brightness profile, often used for the description of
low-luminosity elliptical galaxies and
pseudo-bulges. \citetalias{2011A&A...525A.136B} calculated the
luminosity density of the exponential model by directly deprojecting
the surface brightness profile and through its representation as a
Meijer $G$ function. For the luminosity density we get
\begin{equation}
  \nu(r) 
  = 
  \frac{I_0\,b}{\pi\,\reff}\,
  \sum_{k=0}^{\infty} 
  \frac{1}{k!\,k!}
  \left(\frac{u}{2}\right)^{2k}
  \left[ 
    - \ln\left(\frac{u}{2}\right) + \psi(k+1)
  \right]
  = 
  \frac{I_0\,b}{\pi\,\reff}\,K_0(u),
\label{nu1}
\end{equation}
with $K_\nu(z)$ the modified Bessel function of the second kind. This
expression is in agreement with equation~(24) of
\citetalias{2011A&A...525A.136B}. Similarly, we obtain for the
cumulative luminosity 
%
\begin{multline}
  \hspace*{2cm}
  L(r)
  = 
  \frac{4\,I_0\,\reff^2}{b^2}
  \sum_{k=0}^{\infty} 
  \frac{8}{(2k+3)\,k!\,k!}
  \left(\frac{u}{2}\right)^{2k+3}
  \left[ 
    -\ln\left(\frac{u}{2}\right) + \frac{1}{2k+3} + \psi(k+1)
  \right]
  \\
  =
  \frac{2\pi\,I_0\,\reff^2}{b^2}\,
  u\,\left[K_2(u)\,L_1(u)+K_1(u)\,L_2(u)-\frac{4}{3\pi}\,u\,K_1(u)\right],
  \hspace*{2cm}
\end{multline}
%
with $L_\nu(u)$ the modified Struve function. Finally, for the
potential of the exponential model we get the expansion
%
\begin{multline}
  \hspace*{1cm}
  \Psi(r)
  = 
  \frac{4\,G\,\Upsilon\,I_0\,\reff}{b}
  \left\{
    1
    -
    \sum_{k=0}^{\infty} 
    \frac{2}{(2k+3)\,k!\,(k+1)!}
    \left(\frac{u}{2}\right)^{2k+2}
    \left[ 
      - \ln\left(\frac{u}{2}\right) +  \frac{1}{2k+3} + \frac{1}{2k+2} + \psi(k+1)
    \right]
  \right\}
  \\
  =
  \frac{2\pi\,G\,\Upsilon\,I_0\,\reff}{b}
  \left[K_2(u)\,L_1(u)+K_1(u)\,L_2(u)+\frac{2}{3\pi}\,u\,K_1(u)\right].
  \hspace*{1cm}
\end{multline}
These last two expressions can also be obtained by substituting the
luminosity density~(\ref{nu1}) into the recipes~(\ref{defL}) and
(\ref{defPsi}).

\section{Asymptotic behaviour}
\label{sec:asymptot}

With all the explicit power series of Section~\ref{sec:series}
available, it is fairly straightforward to examine the asymptotic
behaviour of the spatial function of the S\'ersic model at small
radii, generalizing the results of \citet{1991A&A...249...99C} and
\citetalias{2011A&A...525A.136B}. The density has the following rich
behaviour, depending on the value of $m$:
\begin{subequations}
\begin{align}
  \nu(r) &\sim \frac{I_0\,b^m}{\pi\,\reff}
  \left[\Gamma(1-m)\, + \,\frac{1}{2}\,\Gamma(1-3m)\,u^2\right]
  \qquad&&
  \text{for $0<m<\tfrac13$ or $m=\tfrac12$},
  \\[3mm]
  \phantom{\hspace{2cm}}&\phantom{\hspace{10cm}}&&\phantom{\hspace{5cm}}\nonumber\\[-3mm]
  \nu(r) &\sim \frac{I_0\,b^{1/3}}{\pi\,\reff}
  \left[\Gamma\left(\frac{2}{3}\right)\, - \,\frac{1}{2}
    \left(3\ln\left(\frac{u}{2}\right)+\gamma+\frac{3}{2}\right)u^2\right]
  \qquad&&
  \text{for $m = \tfrac13$},
  \\[3mm]
  \nu(r) &\sim \frac{I_0\,b^m}{\pi\,\reff}
  \left[\Gamma(1-m)\, + \frac{\sqrt{\pi}}{2m}\,
    \frac{\Gamma\left(\tfrac12 - 
        \tfrac{1}{2m}\right)}{\Gamma\left(1-\tfrac{1}{2m}\right)}\,u^{1/m-1}\right]
  \qquad&&
  \text{for $\tfrac13<m<1$ and $m\neq\tfrac12$},
  \\[3mm]
  \nu(r) &\sim \frac{I_0\,b}{\pi\,\reff}
  \left[-\ln\left(\frac{u}{2}\right)-\gamma\right]
  \qquad&&
  \text{for $m = 1$},
  \\[3mm]
  \nu(r) &\sim \frac{I_0\,b^m}{\sqrt{\pi}\,\reff}
  \frac{1}{2m}\,
    \frac{\Gamma\left(\tfrac12 - \tfrac{1}{2m}\right)}
    {\Gamma\left(1-\tfrac{1}{2m}\right)}\,u^{1/m-1}
  \qquad&&
  \text{for $m>1$}.
\end{align}
\end{subequations}
with $\gamma \approx 0.57721566$ the Euler-Mascheroni constant. The
luminosity behaves as
\begin{subequations}
\begin{align}
  L(r) &\sim \frac{4\,I_0\,\reff^2}{3\,b^{2m}}\, \Gamma(1-m)\,u^3
  \qquad&&
  \text{for $m < 1$},
  \\[3mm]
  \phantom{\hspace{2cm}}&\phantom{\hspace{10cm}}&&\phantom{\hspace{5cm}}\nonumber\\[-3mm]
  L(r) &\sim \frac{4\,I_0\,\reff^2}{3\,b^2}
  \left[-\ln\left(\frac{u}{2}\right)-\gamma+\frac{1}{3}\right]u^3 
  \qquad&&
  \text{for $m = 1$},
  \\[3mm]
  L(r) &\sim \frac{I_0\,\reff^2}{b^{2m}}\,
  \frac{2\sqrt{\pi}}{2m+1}\,
  \frac{\Gamma\left(\tfrac12 - \tfrac{1}{2m}\right)}
  {\Gamma\left(1-\tfrac{1}{2m}\right)}\,u^{1/m+2}
  \qquad&&
  \text{for $m > 1$}.
\end{align}
\end{subequations}
Finally, the potential approaches $r\rightarrow 0$ as
\begin{subequations}
\begin{align}
  \Psi(r) &\sim \frac{G\,\Upsilon\,I_0\,\reff}{b^m}
  \left[4\,\Gamma(1+m) \,-\, \frac{2}{3}\,\Gamma(1-m)\,u^2\right]
  \qquad&&
  \text{for $m < 1$},
  \\[3mm]
  \phantom{\hspace{2cm}}&\phantom{\hspace{10cm}}&&\phantom{\hspace{5cm}}\nonumber\\[-3mm]
  \Psi(r) &\sim \frac{G\,\Upsilon\,I_0\,\reff}{b}
  \left[4 \,+ \, \frac{2}{3}
    \left(\ln\left(\frac{u}{2}\right)+\gamma-\frac{5}{6}\right)u^2\right]
  \qquad&&
  \text{for $m = 1$},
  \\[3mm]
  \Psi(r) &\sim \frac{G\,\Upsilon\,I_0\,\reff}{b^m}
  \left[4\,\Gamma(1+m)\, + \,\frac{\sqrt{\pi}}{2m+1}\,
    \frac{\Gamma\left(-\tfrac12 - \tfrac{1}{2m}\right)}
    {\Gamma\left(1-\tfrac{1}{2m}\right)}\,u^{1/m+1}\right]
  \qquad&&
  \text{for $m > 1$}.
\end{align}
\label{Psias}
\end{subequations}
If we set $r = 0$ in the expressions~(\ref{Psias}), we recover the
central potential
\begin{equation}
  \Psi_0 = \frac{4\,G\,\Upsilon\,I_0\,\reff}{b^m}\,\Gamma(m+1),
\end{equation}
for every real value $m>0$, in agreement with equation~(12) of
\cite{1991A&A...249...99C}. 

\section{Conclusions}
\label{sec:conclusions}

This paper is a companion paper to \citetalias{2011A&A...525A.136B}
that deals with the derivation of analytical expressions for the
luminosity density $\nu(r)$, cumulative luminosity $L(r)$ and
gravitational potential $\Psi(r)$ of the S\'ersic model. Our work
extends the work initiated by \citet{2002A&A...383..384M}, who managed
to express these properties in terms of the Meijer $G$ function for
integer values of the S\'ersic index $m$. In
\citetalias{2011A&A...525A.136B} we extended this analysis by
demonstrating that these expressions can be derived using a Mellin
integral transform and that they are also valid for half-integer
values of $m$. We also derived more general expressions in terms of
the Meijer $G$ function for these spatial properties for S\'ersic
models with rational $m$. Actually, the Mellin integral transform
approach directly led to an expression in terms of the Fox $H$
function for the luminosity density valid for general values of
$m$. Not fully aware of the rich power of the Fox $H$ function and
attracted by the availability of implementations of the Meijer $G$
function both in symbolic computer algebra packages and as
high-performance computing code, we focused our attention in
\citetalias{2011A&A...525A.136B} on the study of the S\'ersic model
with integer and rational $m$. In this paper, we have extended this
analysis to arbitrary values of $m$, now fully making use of the Fox
$H$ function. We have derived compact and elegant expressions for the
luminosity density, cumulative luminosity and gravitational potential
that are valid for all values of $m$. We have used the properties of
the Fox $H$ function to deduce explicit power and logarithmic-power
expansions for $\nu$, $L$ and $\Psi$ that can be used to evaluate
these important spatial properties for arbitrary values of $m$. These
series expansions also provide a direct way to probe the rich
asymptotic behaviour of the S\'ersic model at small radii; our results
are in full agreement with the more ad hoc analysis of
\citetalias{2011A&A...525A.136B}.

Our work fits into a significant effort of analytical work on the
S\'ersic model: apart from the spatial properties discussed here, also
its photometric \citep{2005PASA...22..118G}, dynamic
\citep{1991A&A...249...99C, 1997A&A...321..724C} and lensing
\citep{2004A&A...415..839C, 2007JCAP...07..006E} properties have
studied extensively through analytical means. Given that the S\'ersic
model is the de facto standard model to describe the surface
brightness distribution of hot stellar systems, this combined range of
analytical properties has plenty of applications. For example, they
can be used for the construction of self-consisting dynamical models
for elliptical galaxies, serve as a realistic starting point for
numerical $N$-body or smoothed particle hydrodynamics simulations, or
serve as a template for radiative transfer or gravitational lens
modelling. We are convinced that the exact analytical results for the
spatial properties of the S\'ersic models will be a useful
contribution in this respect.

Besides this practical use, we hope to have demonstrated the power of
the Fox $H$ as a tool for analytical work. We are aware that the Fox
$H$ function is not the most every day's special function and that its
unfamiliar definition through inverse Mellin transform alone might
already seem daunting. However, the Fox $H$ function is gradually
appearing more in mathematics and applied sciences and there are now
several volumes of comprehensive literature available on its
properties \citep[e.g.][]{MS78, Srivastava82, KilbasSaigo04,
  M+09}. Through this work, we advocate the use of the Fox $H$
function in theoretical astrophysics.

\appendix

\section{The Fox $H$ function}
\label{appendix}

The Fox $H$ function, or just called the $H$ function, is generally
defined as the inverse Mellin transform of a product of gamma
functions,
\begin{equation}
  H^{m,n}_{p,q}\!\left[\left.
      \begin{matrix} 
        ({\boldsymbol{a}},{\boldsymbol{A}})
        \\
        ({\boldsymbol{b}},{\boldsymbol{B}})
      \end{matrix}
      \,\right|\, z \right] 
  = 
  \frac{1}{2\pi i} 
  \int_{\cal{L}}
  \frac{\prod_{j=1}^m \Gamma(b_j+B_js) \prod_{j=1}^n \Gamma(1-a_j-A_js)}
  {\prod_{j=m+1}^q \Gamma(1-b_j-B_js) \prod_{j=n+1}^p\Gamma(a_j+A_js)}\,
  z^{-s}\,{\text{d}}s.
\label{defH}
\end{equation}
It was introduced by \citet{Fox61} in his attempt to find the most
general symmetric Fourier kernel. The Fox $H$ function is a
generalization of the Meijer $G$ function, but includes also many
other special functions, including Mittag-Leffler functions and
generalized Bessel functions. For details on the definition,
convergence and many useful properties of the Fox $H$ function, we
refer to \citet{MS78}, \citet{Srivastava82}, \citet{KilbasSaigo04} or
\citet{M+09}, and the references therein.

We will now derive a series expansion of the Fox $H$ function. Let us
write the Fox $H$ function in the form
\begin{equation}
  H^{m,n}_{p,q}\!\left[\left.
      \begin{matrix} 
        ({\boldsymbol{a}},{\boldsymbol{A}})
        \\
        ({\boldsymbol{b}},{\boldsymbol{B}})
      \end{matrix}
      \,\right|\, z \right] 
  = 
  \frac{1}{2\pi i} 
  \int_{\cal{L}}
  \varphi(s)\,
  z^{-s}\,{\text{d}}s,
\end{equation}
with
\begin{equation}
  \varphi(s) =   
  \frac{\prod_{j=1}^m \Gamma(b_j+B_js) \prod_{j=1}^n \Gamma(1-a_j-A_js)}
  {\prod_{j=m+1}^q \Gamma(1-b_j-B_js) \prod_{j=n+1}^p\Gamma(a_j+A_js)}.
\end{equation}
Under certain conditions (which are always satisfied for the Fox $H$
functions we consider in this paper; see e.g.\ \citet{KilbasSaigo99}
or \citet{M+09} for details), the Fox $H$ function is an analytical
function, and the contour integral can be evaluated using the residue
theorem. If we introduce for the poles of the functions $\Gamma(b_i +
B_is)$ the short-hand notation $\beta_{i,k} = -(b_i + k)/B_i$, then
\begin{equation}
  H^{m,n}_{p,q}\!\left[\left.
      \begin{matrix} 
        ({\boldsymbol{a}},{\boldsymbol{A}})
        \\
        ({\boldsymbol{b}},{\boldsymbol{B}})
      \end{matrix}
      \,\right|\, z \right] 
  = 
  \sum_{i=1}^m\sum_{k=0}^{\infty}\Res_{s=\beta_{i,k}}
  \Big[\varphi(s)\,z^{-s}\Big].
\end{equation}
If $\beta_{i,k}$ is a simple pole, then the corresponding residue is fairly straightforward:
we find
\begin{equation}\label{simplepole}
  \Res_{s=\beta_{i,k}}\Bigl[\varphi(s)\,z^{-s}\Bigr]
  = 
  \lim_{s\rightarrow \beta_{i,k}} (s-\beta_{i,k})\,\varphi(s)\,z^{-s}
  = 
  \varphi_i(\beta_{i,k})\,\frac{(-1)^k}{k!}\,\frac{z^{-\beta_{i,k}}}{B_i},
\end{equation}
where $\varphi_i(s)$ is $\varphi(s)/\Gamma(b_i + B_is)$.  As a result,
if all gamma functions $\Gamma(b_i + B_is)$ have only single poles, we
get as a series expansion,
\begin{equation}
  H^{m,n}_{p,q}\!\left[\left.
      \begin{matrix} 
        ({\boldsymbol{a}},{\boldsymbol{A}})
        \\
        ({\boldsymbol{b}},{\boldsymbol{B}})
      \end{matrix}
      \,\right|\, z \right] 
  = 
  \sum_{i=1}^m
  \sum_{k=0}^\infty
  \frac{(-1)^k}{k!\,B_i}\,
  \frac{\prod_{j=1,j\ne i}^m\Gamma\left(b_j-B_j\frac{b_i+k}{B_i}\right) 
        \prod_{j=1}^n \Gamma\left(1-a_j+A_j\frac{b_i+k}{B_i}\right)}
      {\prod_{j=m+1}^q \Gamma\left(1-b_j+B_j\frac{b_i+k}{B_i}\right) 
        \prod_{j=n+1}^p\Gamma\left(a_j-A_j\frac{b_i+k}{B_i}\right)}\,
      z^{(b_i+k)/B_i},
\label{powerseries}
\end{equation}
in agreement with equation (3.12) in \citet{KilbasSaigo99}. However,
if several gamma functions share the same pole, then this pole is of
second order, and the calculation becomes more cumbersome.
\citet{KilbasSaigo99} demonstrate that the Fox $H$ function can then
be expressed as a logarithmic-power series rather than a simple power
series. They present a generic expression valid for all orders of pole
multiplicity. Here, we present a less general, but more explicit,
series expansion in the case that two gamma functions
$\Gamma(b_j+B_js)$ share at least one pole. Without loss of
generality, we can place these two gamma functions at the front, so
that we can write
\begin{equation}
  \varphi(s) = \Gamma(b_1 + B_1s)\,\Gamma(b_2 + B_2s)\,\phi(s),
\end{equation}
with
\begin{equation}
  \phi(s) = 
  \frac{\prod_{j=3}^m \Gamma(b_j+B_js) \prod_{j=1}^n \Gamma(1-a_j-A_js)}
  {\prod_{j=m+1}^q \Gamma(1-b_j-B_js) \prod_{j=n+1}^p\Gamma(a_j+A_js)}.
\end{equation}
Now, suppose there is a tuple $(k_1,k_2)$ of indices so that
$\beta_{1,k_1} = \beta_{2,k_2}$. The residue of this second-order pole
is then, after some algebra,
\begin{align}
  \Res_{s=\beta_{1,k_1}}\left\{\varphi(s)\,z^{-s}\right\} 
  &= 
  \lim_{s\rightarrow \beta_{1,k_1}}\left\{
    \frac{\text{d}}{\text{d}s}\left[
      (s-\beta_{1,k_1})^2\,\varphi(s)\,z^{-s}\right]\right\}
  \nonumber \\
  &=
  \phi(\beta_{1,k_1})\left[-\ln z + B_1\psi(k_1+1) + B_2\psi(k_2+1) + 
    \frac{\phi'(\beta_{1,k_1})}{\phi(\beta_{1,k_1})}\right]
  \frac{(-1)^{k_1}(-1)^{k_2}}{(k_1)!\,(k_2)!}\,\frac{z^{-\beta_{1,k_1}}}{B_1B_2},
\label{doublepole}
\end{align}
with $\psi(s) = \Gamma'(s)/\Gamma(s)$ the digamma function. Moreover,
note that $\phi(s)$ is a product and quotient of gamma functions, so
that $\phi'(s)/\phi(s)$ can also be expressed as a sum of digamma
functions. This means that all the machinery is available to express
the Fox $H$ function as the rather daunting-looking series expansion
\begin{subequations}
\label{genseries}
\begin{align}
\label{genseriesa}
  H^{m,n}_{p,q}\!\left[\left.
      \begin{matrix} 
        ({\boldsymbol{a}},{\boldsymbol{A}})
        \\
        ({\boldsymbol{b}},{\boldsymbol{B}})
      \end{matrix}
      \,\right|\, z \right] 
  &= 
  \sideset{}{'}\sum_{i,k}
  \frac{(-1)^k}{k!\,B_i}\,
  \frac{\prod_{j=1,j\ne i}^m\Gamma\left(b_j-B_j\frac{b_i+k}{B_i}\right) 
    \prod_{j=1}^n \Gamma\left(1-a_j+A_j\frac{b_i+k}{B_i}\right)}
  {\prod_{j=m+1}^q \Gamma\left(1-b_j+B_j\frac{b_i+k}{B_i}\right) 
    \prod_{j=n+1}^p\Gamma\left(a_j-A_j\frac{b_i+k}{B_i}\right)}\,
 z^{(b_i+k)/B_i}
 \nonumber\\
  &+
  \sideset{}{''}\sum_{k_1}
  \frac{(-1)^{k_1+k_2}}{k_1!\,k_2!\,B_1B_2}\,
  \frac{\prod_{j=3}^m\Gamma\left(b_j-B_j\frac{b_i+k_1}{B_i}\right) 
    \prod_{j=1}^n \Gamma\left(1-a_j+A_j\frac{b_i+k_1}{B_i}\right)}
  {\prod_{j=m+1}^q \Gamma\left(1-b_j+B_j\frac{b_i+k_1}{B_i}\right) 
    \prod_{j=n+1}^p\Gamma\left(a_j-A_j\frac{b_i+k_1}{B_i}\right)}\,
  z^{(b_i+k_1)/B_i}
  \left(C_{k_1}-\ln z\right),
\end{align}
with the constants $C_{k_1}$ defined as
\begin{multline}
\qquad
C_{k_1}
  =
  B_1\psi(k_1+1) 
  + 
  B_2\psi(k_2+1)\ 
  + 
  \sum_{j=3}^{m} B_j\psi\left(b_j-B_j\tfrac{b_1+k_1}{B_1}\right) 
  -
  \sum_{j=1}^n A_j\psi\left(1-a_j+A_j\tfrac{b_1+k_1}{B_1}\right)
  \\
  + 
  \sum_{j=m+1}^q B_j\psi\left(1-b_j+B_j\tfrac{b_1+k_1}{B_1}\right)
  -
  \sum_{j=n+1}^pA_j\psi\left(a_j-A_j\tfrac{b_1+k_1}{B_1}\right).
  \qquad
\end{multline}
\end{subequations}
The prime in the first summation in (\ref{genseriesa}) indicates that
this sum covers only the single poles, and the double prime in the
second summation indicates that this summation runs over the
second-order poles. In the latter summation, we set $k_2 =
B_2(b_1+k_1)/B_1-b_2$.


\begin{thebibliography}{}

\bibitem[Allen et al.(2006)]{2006MNRAS.371....2A} Allen, P.~D., Driver, 
S.~P., Graham, A.~W., Cameron, E., Liske, J., 
\& de Propris, R.\ 2006, MNRAS, 371, 2 

\bibitem[Andredakis et al.(1995)]{1995MNRAS.275..874A} Andredakis,
  Y.~C., Peletier, R.~F., \& Balcells, M.\ 1995, MNRAS, 275, 874

\bibitem[Baes \& Gentile(2011)]{2011A&A...525A.136B} Baes, M., \&
  Gentile, G.\ 2011, A\&A, 525, A136

\bibitem[Baes 
\& van Hese(2007)]{2007A&A...471..419B} Baes, M., \& van Hese, E.\ 2007, \aap, 471, 419 

\bibitem[Caon et al.(1993)]{1993MNRAS.265.1013C} Caon, N., Capaccioli, M., 
 D'Onofrio, M.\ 1993, MNRAS, 265, 1013 

\bibitem[Cardone(2004)]{2004A&A...415..839C} Cardone, V.~F.\ 2004,
  A\&A, 415, 839

\bibitem[Cellone et al.(1994)]{1994ApJS...93..397C} Cellone, S.~A., Forte, 
J.~C., \& Geisler, D.\ 1994, ApJS, 93, 397 

\bibitem[Ciotti(1991)]{1991A&A...249...99C} Ciotti, L.\ 1991, AA, 249, 99 

\bibitem[Ciotti \& Bertin(1999)]{1999A&A...352..447C} Ciotti, L., 
 Bertin, G.\ 1999, A\&A, 352, 447

\bibitem[Ciotti 
\& Lanzoni(1997)]{1997A&A...321..724C} Ciotti, L.,  Lanzoni, B.\ 1997, A\&A, 321, 724 

\bibitem[Davies et al.(1988)]{1988MNRAS.232..239D} Davies, J.~I., 
Phillipps, S., Cawson, M.~G.~M., Disney, M.~J., 
\& Kibblewhite, E.~J.\ 1988, MNRAS, 232, 239 

\bibitem[De Rijcke et al.(2010)]{2010ApJ...724L.171D} De Rijcke, S., Van 
Hese, E., \& Buyle, P.\ 2010, \apjl, 724, L171 

\bibitem[D'Onofrio et al.(1994)]{1994MNRAS.271..523D} D'Onofrio, M., 
Capaccioli, M., \& Caon, N.\ 1994, MNRAS, 271, 523 

\bibitem[El{\'{\i}}asd{\'o}ttir \&
  M{\"o}ller(2007)]{2007JCAP...07..006E} El{\'{\i}}asd{\'o}ttir,
  {\'A}., M{\"o}ller, O.\ 2007, Journal of Cosmology and
  Astro-Particle Physics, 7, 6

\bibitem[Fox(1961)]{Fox61} Fox, C.\ 1961, Transactions of the American
  Mathematical Society, 98, 395

\bibitem[Gadotti(2009)]{2009MNRAS.393.1531G} Gadotti, D.~A.\ 2009, MNRAS, 
393, 1531 

\bibitem[Gradshteyn \& Ryzhik(1965)]{1965tisp.book.....G} Gradshteyn,
  I.~S., \& Ryzhik, I.~M.\ 1965, New York: Academic Press, 1965, 4th
  ed., edited by Geronimus, Yu.V.~(4th ed.); Tseytlin, M.Yu.~(4th
  ed.),

\bibitem[Graham 
\& Driver(2005)]{2005PASA...22..118G} Graham, A.~W.,  Driver, S.~P.\ 2005, Publications of the Astronomical Society of Australia, 22, 118 

\bibitem[Graham 
\& Guzm{\'a}n(2003)]{2003AJ....125.2936G} Graham, A.~W., \&
Guzm{\'a}n, R.\ 2003, AJ, 125, 2936 

\bibitem[Haubold et al.(2007)]{2007BASI...35..681H} Haubold, H.~J.,
  Mathai, A.~M., \& Saxena, R.~K.\ 2007, Bulletin of the Astronomical
  Society of India, 35, 681

\bibitem[Haubold et al.(2011)]{2011arXiv1102.5498H} Haubold, H.~J., Mathai, 
A.~M., \& Saxena, R.~K.\ 2011, arXiv:1102.5498 

\bibitem[Kilbas \& Saigo(1999)]{KilbasSaigo99} Kilbas, A.~A.,  Saigo,
  M.\ 2004, J. Appl. Math. Stochast. Anal., 12, 191

\bibitem[Kilbas \& Saigo(2004)]{KilbasSaigo04} Kilbas, A.~A.,  Saigo,
  M.\ 2004, H-Transforms: Theory and Applications, CRC Press

\bibitem[Mathai \& Saxena(1978)]{MS78} Mathai, A.~M.,  Saxena R.~M.\
 1978, The $H$ Function with Applications in Statistics and Other
 Disciplines, Wiley

\bibitem[Mathai et al.(2009)]{M+09} Mathai, A.~M., Saxena R.~M., 
  Haubold H.~J.\ 2009, The $H$-Function: Theory and Applications,
  Springer

\bibitem[Mazure 
\& Capelato(2002)]{2002A&A...383..384M} Mazure, A.,  Capelato,
H.~V.\ 2002, A\&A, 383, 384 

\bibitem[M{\"o}llenhoff 
\& Heidt(2001)]{2001A&A...368...16M} M{\"o}llenhoff, C., \& Heidt, J.\ 2001, A\&A, 368, 16 

\bibitem[Prugniel \& Simien(1997)]{1997A&A...321..111P} Prugniel, P.,
  \& Simien, F.\ 1997, A\&A, 321, 111

\bibitem[S\'ersic(1968)]{1968adga.book.....S} S\'ersic, J.~L.\ 1968, Cordoba, 
Argentina: Observatorio Astronomico

\bibitem[Srivastava et al.(1982)]{Srivastava82} Srivastava, H.~M.,
  Gupta, K.~C.; Goyal, S.~P.\ 1982, The H-Function of One and Two
  Variables with Applications. New Delhi, India: South Asian Publ.

\bibitem[Trujillo et al.(2001)]{2001MNRAS.326..869T} Trujillo, I., Graham, 
A.~W., \& Caon, N.\ 2001, MNRAS, 326, 869 

\bibitem[Van Hese et al.(2009)]{2009ApJ...690.1280V} Van Hese, E., Baes, 
M., \& Dejonghe, H.\ 2009, \apj, 690, 1280 

\end{thebibliography}
\end{document}